\documentclass[draftcls,onecolumn,12pt]{IEEEtran}

\pdfoutput=1

\usepackage{epsfig}
\usepackage{amsmath}
\usepackage{amsfonts}
\usepackage{amssymb}
\usepackage{subfigure}
\usepackage{booktabs}

\newtheorem{theorem}{Theorem}
\newtheorem{corollary}{Corollary}

\DeclareMathOperator*{\argmax}{arg\,max}

\newcommand{\Xjb}{\Xc_b^j}

\newcommand{\Ccm}[1]{{\sf C}^{\rm cm}_{ #1}}
\newcommand{\Cbicm}[1]{{\sf C}^{\rm bicm}_{ #1}}

\newcommand{\Aeps}{{\mathcal{A}_{\epsilon}}}
\newcommand{\Beps}{{\mathcal{B}_{\epsilon}}}
\newcommand{\Aepsc}{{\mathcal{A}_{\epsilon}^{\rm c}}}
\newcommand{\Bepsc}{{\mathcal{B}_{\epsilon}^{\rm c}}}

\newcommand{\what}{\widehat{\msf}}

\newcommand{\ub}{\bv}
\newcommand{\uy}{\yv}

\newcommand{\prior}{{\rm ext}}
\newcommand{\Prior}{{\rm EXT}}

\newcommand{\Igmi}{I_{\rm gmi}}
\newcommand{\Igmis}{I_{\rm gmi}(s)}

\newcommand{\Mc}{{\cal M}}
\newcommand{\Cc}{{\cal C}}
\newcommand{\msf}{{\sf m}}
\newcommand{\bv}{{\boldsymbol b}}

\newcommand{\xv}{{\boldsymbol x}}
\newcommand{\yv}{{\boldsymbol y}}
\newcommand{\Xc}{{\cal X}}
\newcommand{\Yc}{{\cal Y}}
\newcommand{\eqdef}{\stackrel{\Delta}{=}}
\newcommand{\Xm}{{\boldsymbol X}}
\newcommand{\Ym}{{\boldsymbol Y}}
\newcommand{\CC}{\mathbb{C}}
\newcommand{\EE}{\mathbb{E}}
\newcommand{\snr}{{\sf snr}}
\newcommand{\Csf}{{\sf C}}

\newcommand{\beq}{\begin{equation}}
\newcommand{\eeq}{\end{equation}}
\title{Bit-Interleaved Coded Modulation Revisited: \\A Mismatched Decoding Perspective}

\author{\vspace{-1mm}Alfonso Martinez, Albert Guill\'en i F\`abregas, Giuseppe Caire and Frans Willems
\thanks{A. Martinez is with the Centrum voor Wiskunde en Informatica (CWI), Kruislaan 413, P.O. Box 94079, 1090 GB Amsterdam, The Netherlands, e-mail: {\tt alfonso.martinez@ieee.org}. A. Guill\'en i F\`abregas is with the Department of Engineering, University of Cambridge, Cambridge, CB2 1PZ, UK, e-mail: {\tt guillen@ieee.org}. G. Caire is with the Electrical Engineering Department, University of Southern California, 3740 McClintock Ave., Los Angeles, CA 90080, USA, e-mail: {\tt caire@usc.edu}. F. Willems is with the Department of Electrical Engineering, Technische Universiteit Eindhoven, Postbus 513, 5600 MB Eindhoven, The Netherlands, e-mail: {\tt f.m.j.willems@tue.nl}.}
\thanks{This work has been supported by the International Incoming Short Visits Scheme 2007/R2 of the Royal Society.}
\thanks{This work has been presented in part at the 27th Symposium Information Theory in the Benelux,  wic 2006,  June 2006, Noordwijk, The Netherlands, and at
the IEEE International Symposium on Information Theory, Toronto, Canada, July 2008.}}

\begin{document}

\maketitle

\vspace{-18mm}
\begin{abstract}
We revisit the information-theoretic analysis of bit-interleaved coded modulation (BICM) by modeling the BICM decoder as a mismatched decoder. The mismatched decoding model is well-defined for finite, yet arbitrary, block lengths, and naturally captures the channel memory among the bits belonging to the same symbol. We give two independent proofs of the achievability of the BICM capacity calculated by Caire {\em et al.}  where BICM was modeled as a set of independent parallel binary-input channels whose output is the bitwise log-likelihood ratio. Our first achievability proof uses typical sequences, and shows that due to the random coding construction, the interleaver is not required. The second proof is based on the random coding error exponents with mismatched decoding, where the largest achievable rate is the generalized mutual information.  We show that the generalized mutual information of the mismatched decoder coincides with the infinite-interleaver BICM capacity.  We also show that the error exponent --and hence the cutoff rate-- of the BICM mismatched decoder is upper bounded by that of coded modulation and may thus be lower than in the infinite-interleaved model. For binary reflected Gray mapping in Gaussian channels the loss in error exponent is small. We also consider the mutual information appearing in the analysis of iterative decoding of BICM with EXIT charts. We show that the corresponding symbol metric has knowledge of the transmitted symbol and the EXIT mutual information admits a representation as a pseudo-generalized mutual information, which is in general not achievable. A different symbol decoding metric, for which the extrinsic side information refers to the hypothesized symbol, induces a generalized mutual information lower than the coded modulation capacity. We also show how perfect extrinsic side information turns the error exponent of this mismatched decoder into that of coded modulation.
\end{abstract}

\newpage

\section{Introduction}

In the classical bit-interleaved coded modulation (BICM) scheme proposed by Zehavi in \cite{zehavi1992ptc}, the channel observation is used to generate decoding metrics for each of the bits of a symbol, rather than the symbol metrics used in Ungerb\"ock's coded modulation (CM) \cite{ungerboeck1982ccm}. This decoder is sub-optimal and non-iterative, but offers very good performance and is interesting from a practical perspective due to its low implementation complexity. In parallel, iterative decoders have also received much attention \cite{li1998bic,li1999tcm,tenbrink2000did,tenbrink1998idq,tenbrink1999cid} thanks to their improved performance. 

Caire {\em et al.} \cite{caire1998bic} further elaborated on Zehavi's decoder and, under the assumption of an infinite-length interleaver, presented and analyzed a BICM channel model as a set of parallel independent binary-input output symmetric channels.
Based on the data processing theorem \cite{cover1991eit}, Caire {\em et al.} showed that the BICM mutual information cannot be larger than that of CM. However, and rather surprisingly a priori, they found that the cutoff rate of BICM might exceed that of CM \cite{wozencraft1965pce}. The error exponents for the parallel-channel model were studied by Wachsmann {\em et al.} \cite{wachsmann1999mct}.

In this paper we take a closer look to the classical BICM decoder proposed by Zehavi and cast it as a mismatched decoder \cite{merhav1994irm,kaplan1993ira,ganti2000mdr}. The observation that the classical BICM decoder treats the different bits in a given symbol as independent, even if they are clearly not, naturally leads to a simple model of the symbol mismatched decoding metric as the product of bit decoding metrics, which are in turn related to the log-likelihood ratios. 
We also examine the BICM mutual information in the analysis of iterative decoding by means of EXIT charts \cite{tenbrink2000did,tenbrink1998idq,tenbrink1999cid}, where the sum of the mutual informations across the parallel subchannels is used as a figure of merit of the progress in the iterative decoding process. 

This paper is organized as follows. Section~\ref{section:model} introduces the system model and notation. Section~\ref{section:achievability} gives a proof of achievability of the BICM capacity, derived in \cite{caire1998bic} for the independent parallel channel model, by using typical sequences.
Section~\ref{section:randomCodes} shows general results on the error exponents, including the generalized mutual information and cutoff rate as particular instances. 
The BICM error exponent (and in particular the cutoff rate) is always upper-bounded by that of CM, as opposed to the corresponding exponent for the independent parallel channel model \cite{caire1998bic,wachsmann1999mct}, which can sometimes be larger.
In particular, Section~\ref{section:achievability-GMI} studies the achievable rates of BICM under mismatched decoding and shows that the generalized mutual information \cite{merhav1994irm,kaplan1993ira,ganti2000mdr} of the BICM mismatched decoder yields the BICM capacity.
The section concludes with some numerical results, including a comparison with the parallel-channel models. In general, the loss in error exponent is negligible for binary reflected Gray mapping in Gaussian channels.
In Section~\ref{section:feedback} we turn our attention to the iterative decoding of BICM. First, we review how the mutual information appearing in the analysis of iterative decoding of BICM with EXIT charts, where the symbol decoding metric has some side knowledge of the transmitted symbol, admits a representation as a pseudo-generalized mutual information. A different symbol decoding metric, for which the extrinsic side information refers to the hypothesized symbol, induces a generalized mutual information lower in general than the coded modulation capacity. Moreover, perfect extrinsic side information turns the error exponent of this mismatched decoder into that of coded modulation.
Finally, Section~\ref{section:conclusions} draws some concluding remarks.

\section{Channel Model and Code Ensembles}
\label{section:model}

\subsection{Channel Model}
We consider the transmission of information by means of a block code $\Mc$ of length $N$. 
At the transmitter, a message $\msf$ is mapped onto a codeword $\xv=(x_1,\dotsc,x_N)$, according to one of the design options described later, in Section~\ref{section:ensembles}. We denote this encoding function by $\phi$.
Each of the symbols $x$ are drawn from a discrete modulation alphabet $\Xc = \{x_1,\dotsc,x_M\}$, with $M\eqdef|\Xc|$ and $m = \log_2M$ being the number of bits required to index a symbol. 

We denote the output alphabet by $\Yc$ and the channel output by $\yv\eqdef(y_1,\dotsc,y_N)$, with $y_k\in\Yc$. With no loss of generality, we assume the output is continuous\footnote{All our results are directly applicable to discrete output alphabets, by appropriately replacing integrals by sums.}, so that the channel output $\yv$ related to the codeword $\xv$ through a conditional probability density function $p(\yv|\xv)$. 
Further, we consider memoryless channels, for which
\beq
p(\yv|\xv) = \prod_{k=1}^N p(y_k|x_k),
\eeq
where $p(y|x)$ is the channel symbol transition probability. Henceforth, we drop the words density function in our references of $p(y|x)$.
We denote by $X,Y$ the underlying random variables. 
Similarly, the corresponding random vectors are
$\Xm \eqdef (\underbrace{X,\dotsc,X}_{N~\text{times}})$ and $\Ym \eqdef (\underbrace{X,\dotsc,X}_{N~\text{times}})$,
respectively drawn from the sets
$\boldsymbol{\Xc}\eqdef \Xc^N$, $\boldsymbol{\Yc}\eqdef \Yc^N$.

A particularly interesting, yet simple, case is that of complex-plane signal sets in AWGN with fully-interleaved fading where $\Yc=\CC$ and
\beq
y_k=h_k\sqrt{\snr} \,x_k +z_k,~~~~ k=1,\dotsc,N
\label{eq:model}
\eeq 
where $h_k$ are fading coefficients with average unit energy, $z_k$ are the complex zero-mean unit-variance AWGN samples and $\snr$ is the signal-to-noise ratio (SNR). 
The decoder outputs an estimate of the message $\widehat{\msf}$ according to a given codeword {\em decoding metric}, which we denote by $q(\xv,\yv)$ as
\begin{align}
\widehat{\msf}& = \argmax_{\msf} q(\xv_\msf,\yv) 
\label{eq:min_metric_dec}.
\end{align}
The codeword metrics we consider are the product of symbol decoding metrics $q(x,y)$, for $x\in\Xc$ and $y\in\Yc$, namely
\beq
q(\xv,\yv) = \prod_{k=1}^Nq(x_k,y_k).
\label{eq:symbol_metric}
\eeq 

Assuming that the codewords have equal probability, this decoder finds the most likely codeword as long as $q(x,y)= f\bigl( p(y|x)\bigr)$, where $f(.)$ is a one-to-one increasing function, i.e., as long as the decoding metric is a one-to-one increasing mapping of the transition probability of the  memoryless channel. If the decoding metric $q(x,y)$ is not an increasing one-to-one function of the channel transition probability, the decoder defined by \eqref{eq:min_metric_dec} is a {\em mismatched decoder} \cite{merhav1994irm,kaplan1993ira,ganti2000mdr}. 

\subsection{Code Ensembles}
\label{section:ensembles}

\subsubsection{Coded Modulation}

In a {\em coded modulation} (CM) scheme $\Mc$, the encoder $\phi$ selects a codeword of $N$ modulation symbols, $\xv_\msf = (x_1,\dotsc,x_N)$ according to the information message $\msf$.
The code is in general non-binary, as symbols are chosen according to a probability law $p(x)$.  
Representing the information message set $\{1,\dotsc,|\Mc|\}$, we have that the rate $R$ of this scheme in bits per channel use is 
given by $R = \frac{K}{N}$, where $K\eqdef \log_2|\Mc|$ denotes the number of bits needed to represent every information message.
At the receiver, a maximum metric decoder $\phi$ (as in Eq.~\eqref{eq:min_metric_dec}) acts on the received sequence $\yv$ to generate an estimate of the transmitted message, $\varphi(\yv) = \widehat{\msf}$.
In coded modulation constructions, such as Ungerboeck's \cite{ungerboeck1982ccm}, the symbol decoding metric is proportional to the channel transition probability, that is $q(x,y)\propto p(y|x)$; the value of proportionality constant is irrelevant, as long as it is not zero.
Reliable communication is possible at rates lower than {\em coded modulation capacity} or CM capacity, denoted  by $\Ccm{\Xc}$ and given by
\begin{align}
\Ccm{\Xc} \eqdef \EE\Biggl[\log \frac{ p(Y|X)}{{\sum_{x'\in\Xc} p(x')p(Y|x')}}\Biggr].
\label{eq:c_cm}
\end{align}
The expectation is done according to $p(x)p(y|x)$. We consider often a uniform input distribution $p(x) = 2^{-m}$.

\subsubsection{Bit-Interleaved Coded Modulation}
\label{section:bicm_model_ch}
In a {\em bit-interleaved coded modulation} scheme $\Mc$, the encoder is restricted to be the serial concatenation of a binary code $\Cc$ of length $n \eqdef  mN$ and rate $r=\frac{R}{m}$, a bit interleaver, and a binary labeling function $\mu:\{0,1\}^m \to \Xc$ which maps blocks of $m$ bits to signal constellation symbols. The codewords of $\Cc$ are denoted by $\bv = (b_1,\dotsc,b_{mN})$. The portions of codeword allocated to the $j$-th bit of the label are denoted by $\bv_j\eqdef(b_j,b_{m+j},\dotsc,b_{m(N-1)+j})$. 
We denote the inverse mapping function for labeling position $j$ as $b_j: \Xc\to \{0,1\}$, that is, $b_j(x)$ is the $j$-th bit of symbol $x$. Accordingly, we now define the sets $\Xjb\eqdef\{x\in\Xc : b_j(x)=b\}$
as the set of signal constellation points $x$ whose binary label has value $b\in\{0,1\}$ in its $j$-th position. With some abuse of notation, we will denote $B_1,\dotsc,B_m$ and $b_1,\dotsc,b_m$ the random variables and their corresponding realizations of the bits in a given label position $j=1,\dotsc,m$.

The classical BICM decoder \cite{zehavi1992ptc} treats each of the $m$ bits in a symbol as independent and uses a symbol decoding metric proportional to the product of the a posteriori marginals $p(b_j = b | y)$. More specifically, we have that
\beq
q(x,y)= \prod_{j=1}^m q_j\bigl(b_j(x),y\bigr),
\label{eq:bit_metric_bicm}
\eeq
where the $j$-th bit decoding metric $q_j(b,y)$ is given by
\beq
q_j\bigl(b_j(x)=b,y\bigr) = \sum_{x'\in\Xjb} p(y|x').
\label{eq:bit_metric_sum}
\eeq
This metric is proportional to the transition probability of the output $y$ given the bit $b$ at position $j$, which we denote for later use by $p_j(y|b)$, 
\beq
p_j(y|b) \eqdef \frac{1}{\bigl|\Xjb\bigr|}\sum_{x'\in\Xjb} p(y|x').
\label{eq:bit_marginal}
\eeq

\begin{figure}[htb]
  \centering
  \includegraphics{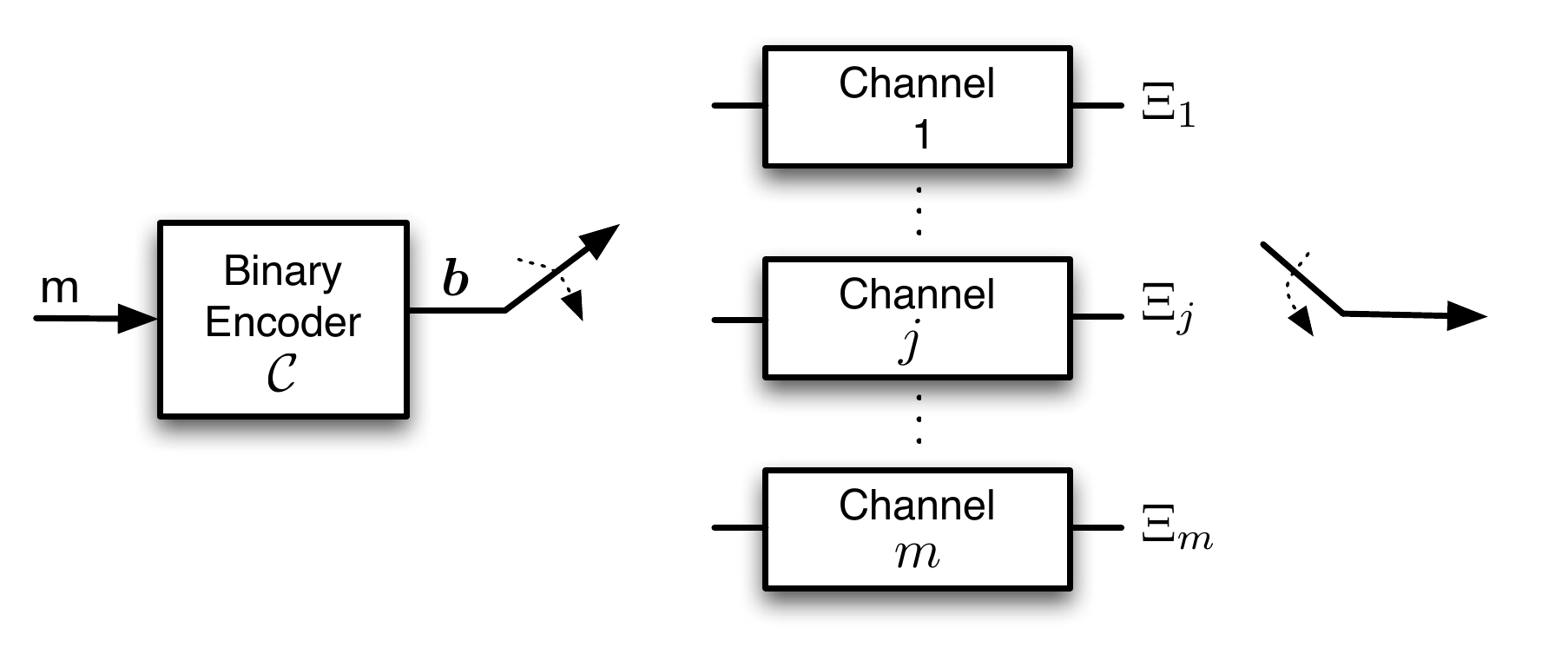}
  \caption{Parallel channel model of BICM.}
  \label{fig:bicm_parallel_channels}
\end{figure}

The set of $m$ probabilities $p_j(y|b)$ can be used as departure point to define an equivalent BICM channel model. Accordingly, Caire {\em et al.} defined a BICM channel \cite{caire1998bic} as the set of $m$ parallel channels having bit $b_j(x_k)$ as input and the bit log-metric (log-likelihood) ratio for the $k$-th symbol
\beq
\Xi_{m(k-1) +j} = \log\frac{q_j\bigl(b_j(x_k)=1,y\bigr)}{q_j\bigl(b_j(x_k)=0,y\bigr)}
\eeq
as output, for $j=1,\dotsc,m$ and $k=1,\dotsc,N$. This channel model is schematically depicted in Figure~\ref{fig:bicm_parallel_channels}. With infinite-length interleaving, the $m$ parallel channels were assumed to be independent in \cite{caire1998bic,wachsmann1999mct}, or in other words, the correlations among the different subchannels are neglected. For this model, Caire {\em et al.} defined a {\em BICM capacity} $\Cbicm{\Xc}$, given by
\begin{align}
\Cbicm{\Xc} &\eqdef \sum_{j=1}^mI(B_j;Y)= \sum_{j=1}^m \EE \Biggl[\log \frac{{\sum_{x'\in\Xc_{B}^j} p(Y|x')}}{\frac{1}{2}\sum_{x'\in\Xc} p(Y|x')}\Biggr],
\end{align}
where the expectation is taken according to $p_j(y|b)p(b)$, for $b\in\{0,1\}$ and $p(b) = \frac{1}{2}$.

In practice, due to complexity limitations, one might be interested in the following lower-complexity version of \eqref{eq:bit_metric_sum},
\beq
q_j(b,y) = \max_{x\in\Xjb} p(y|x).
\label{eq:max_metric}
\eeq
In the log-domain this is known as the max-log approximation. 

Summarizing, the decoder of $\Cc$ uses a mismatched metric of the form given in Eq.~\eqref{eq:symbol_metric} where the decoder of $\Cc$ outputs a binary codeword $\hat\bv$ according to
\beq
\hat\bv =\argmax_{\bv\in\Cc} \prod_{k=1}^N\prod_{j=1}^m q_j\bigl(b_j(x_k),y_n\bigr).
\eeq

\section{Achievability of the BICM Capacity: Typical Sequences}
\label{section:achievability}

In this section, we provide an achievability proof for the BICM capacity based on typical sequences. The proof is based on the usual random coding arguments \cite{cover1991eit} with typical sequences, with a slight modification to account for the mismatched decoding metric. This result is obtained without recurring to an infinite interleaver to remove the correlation among the parallel subchannels of the classical BICM model. We first introduce some notation needed for the proof.

We say that a rate $R$ is achievable if, for every $\epsilon>0$ and for $N$ sufficiently large, there exists an encoder, a demapper and a decoder such that $\tfrac{1}{N}\log |\Mc| \geq R-\epsilon$ and $\Pr(\widehat{\msf}\neq \msf) \leq \epsilon$.
We define the joint probability of the channel output $y$ and the corresponding input bits $(b_{1},\dotsc,b_{m})$ as
\begin{equation}
    p(b_1,\dotsc,b_m,y) \eqdef \Pr\bigl(B_{1}=b_1,\dotsc,B_{m}=b_m,y \leq Y <  y+dy \bigr),
\end{equation}
for all $b_j\in\{0,1\}$, $y$ and infinitely small $dy$.
We denote the derived marginals by $p_j(b_j)$, for $j = 1,\dotsc,m$, and $p(y)$.
The marginal distributions with respect to bit $B_j$ and $Y$ are special, and are denoted by $p_j(b_j,y)$.
We have then the following theorem.
\begin{theorem}
The BICM capacity $\Cbicm{\Xc}$ is achievable.
\end{theorem}

\begin{proof}
Fix an $\epsilon>0$. For each $\msf\in\Mc$ we generate a binary codeword $\ub_1(\msf)\dotsc,\ub_m(\msf)$ with probabilities $p_j(\ub_j)$. The codebook is the set of all codewords generated with this method.

We consider a threshold decoder, which outputs $\what$ only if $\what$ is the unique integer satisfying
\begin{equation}
\bigl(\ub_1(\what),\dotsc,\ub_m(\what),\uy\bigr) \in \Beps,
\end{equation}
where $\Beps$ is a set defined as
\begin{equation}\label{eq:beps}
\Beps \eqdef \Biggl\{ \bigl(\ub_1,\dotsc,\ub_m,\uy\bigr) : \frac{1}{N}\sum_{j=1}^m\log\frac{p_j\bigl(\ub_j(\what),\uy\bigr)}{p_j(\ub_j)\, p(\uy)} \geq \Delta_\epsilon \Biggr\}
\end{equation}
for $\Delta_\epsilon \eqdef \sum_{j=1}^m I(B_j;Y) - 3m\epsilon$. Otherwise, the decoder outputs an error flag.

The usual random coding argument \cite{cover1991eit} shows that the error probability, averaged over the ensemble of randomly generated codes, $\bar{P_e}$, is upper bounded by
\begin{align}
    \bar{P_e} \leq P_1 + (|\Mc|-1)P_2,
\end{align}
where $P_1$ is the probability that the received $\uy$ does not belong to the set $\Beps$,
\begin{equation}\label{eq:pi1}
P_1 \eqdef \sum_{(\ub_1,\dotsc,\ub_m,\uy) \notin \Beps} p(\ub_1,\dotsc,\ub_m,\uy),
\end{equation}
and $P_2$ is the probability that another randomly chosen codeword would be (wrongly) decoded, that is,
\begin{equation}\label{eq:pi2}
P_2 \eqdef \sum_{(\ub_1,\dotsc,\ub_m,\uy) \in \Beps} p(\uy)\prod_{j=1}^m p_j(\ub_j).
\end{equation}

First, we prove that $\Beps \supseteq \Aeps(B_1,\dotsc,B_m,Y)$, where $\Aeps$ is the corresponding jointly typical set \cite{cover1991eit}.
By definition, the sequences $\bigl(\ub_1,\dotsc,\ub_m,\uy\bigr)$ in the typical set satisfy (among other constraints) the following
\begin{align}
&-\log p(\uy) > N(H(Y)-\epsilon), \\
&-\log p_j(\ub_j) > N(H(B_j)-\epsilon), \quad j = 1,\dotsc,m, \\
&-\log p\bigl(\ub_j,\uy\bigr) < N(H(B_j,Y)+\epsilon),\quad j = 1,\dotsc,m.
\end{align}
Here $H(\cdot)$ are the entropies of the corresponding random variables. Multiplying the last equation by $(-1)$, and summing them, we have
\begin{align}
    \log\frac{p_j\bigl(\ub_j(\what),\uy\bigr)}{p_j(\ub_j)p(\uy)} &\geq N\bigl(H(B_j)+H(Y)-H(B_j,Y)-3\epsilon\bigr) \\
    &= N\bigr(I(B_j;Y)-3\epsilon\bigr),
\end{align}
where $I(B_j;Y)$ is the corresponding mutual information.
Now, summing over $j = 1,\dotsc,m$ we obtain
\begin{equation}
    \sum_{j=1}^m\log\frac{p_j\bigl(\ub_j(\what),\uy\bigr)}{p_j(\ub_j)p(\uy)} \geq N\biggl(\sum_{j=1}^m I(B_j;Y)-3\epsilon\biggr) = N\Delta_\epsilon.
\end{equation}
Hence, all typical sequences belong to the set $\Beps$, that is, $\Aeps \subseteq \Beps$. This implies that $\Bepsc \subseteq \Aepsc$ and, therefore, the probability $P_1$ in Eq.~\eqref{eq:pi1} can be upper bounded as
\begin{align}
P_1
&\leq \sum_{(\ub_1,\dotsc,\ub_m,\uy) \notin \Aeps} p(\ub_1,\dotsc,\ub_m,\uy)  \nonumber \\
&\leq \epsilon,
\end{align}
for $N$ sufficiently large. The last inequality follows from the definition of the typical set.

We now move on to $P_2$.
For $(\ub_1,\dotsc,\ub_m,\uy) \in \Beps$, and from the definition of $\Beps$, we have that
\begin{equation}
    2^{N\Delta_{\epsilon}} \leq \prod_{j=1}^m \frac{p_j\bigl(\ub_j,\uy\bigr)}{p_j(\ub_j)p(\uy)} = \prod_{j=1}^m \frac{p_j(\ub_j|\uy)}{p_j(\ub_j)}.
\end{equation}
Rearranging terms we have
\begin{equation}
\prod_{j=1}^m p_j(\ub_j) \leq  \frac{1}{2^{N\Delta_{\epsilon}}} \prod_{j=1}^m p_j(\ub_j|\uy).
\end{equation}
Therefore the probability $P_2$ in Eq.~\eqref{eq:pi2} can be upper bounded
\begin{align}
P_2 
&\leq  \frac{1}{2^{N\Delta_{\epsilon}}} \sum_{(\ub_1,\dotsc,\ub_m,\uy) \in \Beps} p(\uy) \prod_{j=1}^m p_j(\ub_j|\uy)  \\
&\leq  \frac{1}{2^{N\Delta_{\epsilon}}} \sum_{(\ub_1,\dotsc,\ub_m,\uy)} p(\uy) \prod_{j=1}^m p_j(\ub_j|\uy)  \\
&=  \frac{1}{2^{N\Delta_{\epsilon}}} \sum_{(\ub_1,\dotsc,\ub_m,\uy)} p(\ub_1,\dotsc,\ub_m,\uy) \\
&=  \frac{1}{2^{N\Delta_{\epsilon}}}.
\end{align}

Now we can write for $\bar{P_e}$,
\begin{align}
\bar{P_e}
&\leq P_1 + (|\Mc|-1)P_2 \nonumber \\
&\leq \epsilon + |\Mc| 2^{-N\Delta_{\epsilon}} \nonumber \\
&\leq 2\epsilon,
\end{align}
for $|\Mc| = 2^{N(\Delta_{\epsilon}-\epsilon)}$ and large enough $N$.
We conclude that for large enough $N$ there exist codes such that
\begin{align}
    \frac{1}{N} \log_2 |\Mc| \geq \Delta_{\epsilon}-\epsilon = \sum_{j=1}^m I(B_j;Y) -(3m+1)\epsilon, 
\end{align}
and $\Pr\bigl( \widehat{\msf} \neq \msf \bigr) \leq \epsilon$. The rate $\sum_{j=1}^m I(B_j;Y)$ is thus achievable.

To conclude, we verify that the BICM decoder is able to determine the probabilities required for the decoding rule defining $\Beps$ in Eq.~\eqref{eq:beps}. Since the BICM decoder uses the metric $q_j(b_j,y)\propto p_j(y|b_j)$, the log-metric-ratio , or equivalently the a posteriori bit probabilities $p_j(b_j|y)$,
it can also compute
\begin{align}
    \frac{p_j(b_j,y)}{p_j(b_j)\,p(y)} &= \frac{p_j(b_j|y)}{p_j(b_j)},
\end{align}
where the bit probabilities are known, $p_j(1) = p_j(0) = \tfrac{1}{2}$.
\end{proof}

\section{Achievability of the BICM Capacity: Error Exponents, Generalized Mutual Information and Cut-off Rate}
\label{section:randomCodes}

\subsection{Random coding exponent}

The behaviour of the average error probability of a family of randomly generates, decoded with a maximum-likelihood decoder, i.\ e.\ for a decoding metric satisfying $q(x,y) = p(y|x)$, was studied by Gallager in \cite{gallager1968ita}. In particular, Gallager showed the error probability decreases exponentially with the block length $N$ according to a parameter called the {\em error exponent}, provided that the code rate $R$ is below the channel capacity $\Csf$.

For memoryless channels Gallager found \cite{gallager1968ita} that the average error probability over the random coding ensemble can be bounded as
\begin{align}
\bar{P_e} &\leq \exp\Bigl(-N\bigl(E_0(\rho) - \rho R \bigr)\Bigr)
\end{align}
where $E_0(\rho)$ is the Gallager function, given by
\beq\label{eq:E0rho}
E_0(\rho) \eqdef -\log \left(\int_y \left(\sum_{x} p(x) p(y|x)^\frac{1}{1+\rho}\right)^{1+\rho}\,dy\right),
\eeq
and $0\leq \rho\leq 1$ is a free parameter. For a particular input distribution $p(x)$, the tightest bound is obtained by optimizing over $\rho$, which determines the {\em random coding exponent}
\beq
E_{\rm r}(R) = \max_{0\leq\rho\leq1} \, E_0(\rho) - \rho R.
\eeq

For uniform input distribution, we define the {\em coded modulation exponent} $E_0^{\rm cm}(\rho)$ as the exponent of a decoder which uses metrics $q(x,y)= p(y|x)$, namely
\begin{align}
E_0^{\rm cm}(\rho)&= -\log \EE \left[\left(\frac{1}{2^m}\sum_{x'} \left(\frac{p(Y|x')}{p(Y|X)}\right)^\frac{1}{1+\rho}\right)^\rho\right].
\label{eq:exp_cm}
\end{align}

Gallager's derivation can easily be extended to memoryless channels with generic codeword metrics decomposable as product of symbols metrics, that is $q(\xv,\yv) = \prod_{n=1}^Nq(x_n,y_n)$,
Details can be found in \cite{kaplan1993ira}. The error probability is upper bounded by the expression
\begin{align}
    \bar{P_e} &\leq \exp\Bigl(-N\bigl(E_0^q(\rho,s) - \rho R \bigr)\Bigr),
\end{align}
where

the generalized Gallager function $E_0^q(\rho,s)$ is given by
\beq\label{eq:E0rhos}
E_0^q(\rho,s) = - \log \EE\left[\left( \sum_{x'}p(x')\frac{q(x',Y)^{s}}{q(X,Y)^{s}}\right)^\rho \right].
\eeq
The expectation is carried out according to the joint probability $p(y|x)p(x)$.
For a particular input distribution $p(X)$, the random coding error exponent $E^{q}_{\rm r}(R)$ is then given by \cite{kaplan1993ira}
\beq
E^{q}_{\rm r}(R) = \max_{0\leq \rho \leq 1} \max_{s>0}E^q_0(\rho,s) - \rho R.
\eeq

For the specific case of BICM, assuming uniformly distributed inputs and a generic bit metric $q_j(b,y)$, we have that Gallager's generalized function $E^{\rm bicm}_0(\rho,s)$ is given by
\begin{align}\label{eq:E0bicm}
E^{\rm bicm}_0(\rho,s)
&= - \log \EE\left[\left( \frac{1}{2^m}\sum_{x'}\prod_{j=1}^m\frac{ q_j(b_j(x'),Y)^{s}}{ q_j(b_j(X),Y)^{s}}\right)^\rho \right].
\end{align}

For completeness, we note that the cutoff rate is given by $R_0 = E_0(1)$ and, analogously, we define the generalized cutoff rate as
\beq
R_0^q \eqdef E_{\rm r}^q (R=0) =   \max_{s>0} \, E_0^q(1,s).
\eeq

\subsection{Data processing inequality for error exponents}
\label{sec:dataProcessing}

In \cite{kaplan1993ira}, it was proved that the data-processing inequality holds for error exponents, in the sense that for a given input distribution we have that
$E_0^q(\rho,s) \leq E_0(\rho)$ for any $s > 0$.
Next, we rederive this result by extending Gallager's reasoning in \cite{gallager1968ita} to mismatched decoding.

The generalized Gallager function $E^q_0(\rho,s)$ in Eq.~\eqref{eq:E0rhos} can be expressed as
\begin{align}
E_0^q(\rho,s) &= -\log \left(\int_y \sum_{x}p(x)p(y|x) \left(\sum_{x'} p(x') \left(\frac{q(x',y)}{q(x,y)}\right)^s\right)^\rho\right) d y.
\label{eq:expomemt_dpi}
\end{align}
As long as the metric does not depend on the transmitted symbol $x$, the function inside the logarithm can be rewritten as
\begin{align}
 \int_{y}\left(\sum_{x}p(x)p(y|x)q(x,y)^{-s\rho}\right) \left(\sum_{x'} p(x') q(x',y)^s\right)^\rho d y.
 \label{eq:inside_log_hoelder}
\end{align}

For a fixed channel observation $y$, the integrand
is reminiscent of the right-hand side of H{\"o}lder's inequality (see Exercise 4.15 of \cite{gallager1968ita}), which can be expressed as
\begin{align}
    \biggl(\sum_i a_ib_i\biggr)^{1+\rho} \leq \biggl(\sum_i a_i^{1+\rho}\biggr)\biggl(\sum_i b_i^\frac{1+\rho}{\rho}\biggr)^\rho.
\end{align}
Hence, with the identifications
\begin{align}
    a_i &= p(x)^\frac{1}{1+\rho}p(y|x)^\frac{1}{1+\rho}q(x,y)^\frac{-s\rho}{1+\rho}\\
    b_i &= p(x)^\frac{\rho}{1+\rho}q(x,y)^\frac{s\rho}{1+\rho},
\end{align}
we can lower bound Eq.~\eqref{eq:inside_log_hoelder} by the quantity
\begin{align}
    \int_{y}\left(\sum_{x}p(x)p(y|x)^\frac{1}{1+\rho}d y\right)^{1+\rho}.\label{eq:lowerBound-final}
\end{align}

Recovering the logarithm in Eq.~\eqref{eq:expomemt_dpi}, for a general mismatched decoder, arbitrary $s>0$ and any input distribution, we obtain that 
\beq
E_0(\rho) \geq E_0^q(\rho,s).
\label{eq:data_processing_err_exp}
\eeq

Note that the expression in Eq.~\eqref{eq:lowerBound-final} is independent of $s$ and of the specific decoding metric $q(x,y)$. Nevertheless, evaluation of Gallager's generalized function for the specific choices $s = \frac{1}{1+\rho}$ and $q(x,y) \propto p(y|x)$ attains the lower bound, which is also Eq.~\eqref{eq:exp_cm}. 

Equality holds in H{\"o}lder's inequality if and only if for all $i$ and some positive constant $c$, $a_i^\frac{\rho}{1+\rho} = c\, b_i^\frac{1}{1+\rho}$
(see Exercise 4.15 of \cite{gallager1968ita}). In our context, simple algebraic manipulations show that
the necessary condition for equality to hold is that
\begin{equation}\label{eq:sufficiencyHoelder}
    p(y|x) = c' q(x,y)^{s'} \qquad \text{for all $x\in\Xc$}
\end{equation}
for some constants $c'$ and $s'$.
In other words, the metric $q(x,y)$ must be proportional to a power of the channel transition probability $p(y|x)$, for the bound \eqref{eq:data_processing_err_exp} to be tight, and therefore, to achieve the coded modulation error exponent.

\subsection{Error exponent for BICM with the parallel-channel model}
\label{sec:exponent-BICM}

In their analysis of multilevel coding and successive decoding, Wachsmann {\em et al.} provided the error exponents of BICM modelled as a set of parallel channels \cite{wachsmann1999mct}.
The corresponding Gallager's function, which we denote by $E_0^{\rm ind}(\rho)$, is given by
\begin{align}
E_0^{\rm ind}(\rho) 
 &= - \sum_{j=1}^m\log \int_{y}\sum_{b_j=0}^1 p_j(b_j)p_j(y|b_j)\left( \sum_{b_j'=0}^1p_j(b_j')\frac{q_j(b_j',y)^{\frac{1}{1+\rho}}}{q_j(b_j,y)^{\frac{1}{1+\rho}}}\right)^\rho dy , 
 \label{eq:eo_bicm_indep}
\end{align}
which corresponds to a binary-input channel with input $b_j$, output $y$ and bit metric matched to the transition probability $p_j(y|b_j)$.

This equation can be rearranged into a form similar to the one given in previous sections. First, we insert the summation in the logarithm,
\begin{align}
E_0^{\rm ind}(\rho) &= - \log \left(\prod_{j=1}^m\int_{y}\sum_{b_j=0}^1 p_j(b_j)p_j(y|b_j)\left( \sum_{b_j'=0}^1p_j(b_j')\frac{q_j(b_j',y)^{\frac{1}{1+\rho}}}{q_j(b_j,y)^{\frac{1}{1+\rho}}}\right)^\rho dy\right).
\end{align}
Then, we notice that the output variables $y$ are dummy variables which possibly vary for each value of $j$. Let us denote the dummy variable in the $j$-th subchannel by $y_j'$. We have then
\begin{align}
E_0^{\rm ind}(\rho) &= - \log \left(\prod_{j=1}^m\int_{y_j'}\sum_{b_j=0}^1 p_j(b_j)p(y_j'|b_j)\left( \sum_{b_j'=0}^1p_j(b_j')\frac{q_j(b_j',y_j')^{\frac{1}{1+\rho}}}{q_j(b_j,y_j')^{\frac{1}{1+\rho}}}\right)^\rho dy_j' \right) \\
&= - \log \left(\int_{\yv'}\sum_{x} p(x)p(\yv'|x)\left( \sum_{x'}p(x')\frac{q(x',\yv')^{\frac{1}{1+\rho}}}{q(x,\yv')^{\frac{1}{1+\rho}}}\right)^\rho d\yv'\right).
\end{align}
Here we carried out the multiplications, defined the vector $\yv'$ to be the collection of the $m$ channel outputs, and denoted by $x=\mu(b_1,\dotsc,b_m)$ and $x'=\mu(b_1',\dotsc,b_m')$ the symbols selected by the bit sequences.
This equation is the Gallager function of a mismatched decoder for a channel output $\yv'$, such that for each of the $m$ subchannels sees a statistically independent channel realization from the others.

In general, since the original channel cannot be decomposed into parallel, conditionally independent subchannels, this parallel-channel model fails to capture the statistics of the channel. 

The cut-off rate with the parallel-channel model is given by
\begin{align}
R_0^{\rm ind} &= - \sum_{j=1}^m\log \int_{y}\left(\sum_{b_j=0}^1 p_j(b_j)p_j(y|b_j)^{\frac{1}{2}}\right)^2 dy.
\end{align}
The cutoff rate was given in \cite{caire1998bic} as $m$ times the cutoff rate of an averaged channel,
\begin{align}
R_0^{\rm av}     &\eqdef m\left[\log2-\log\left(1+\frac{1}{m}\sum_{j=1}^m\EE \left[\sqrt{\frac{\sum_{x'\in\Xc_{\bar B}^j} p(Y|x')}{\sum_{x'\in\Xc_B^j} p(Y|x')}}\right] \right)\right].
\end{align}
From Jensen's inequality one easily obtains that $R_0^{\rm av} \leq R_0^{\rm ind}$.

\subsection{Generalized mutual information for BICM}
\label{section:achievability-GMI}

The largest achievable rate with mismatched decoding is not known in general. Perhaps the easiest candidate to deal with is the generalized mutual information (GMI) \cite{merhav1994irm,kaplan1993ira,ganti2000mdr}, given by
\begin{align}
\Igmi &\eqdef \sup_{s>0}\Igmis,
\end{align}
where
\begin{align}\label{eq:gmi-s}
\Igmis
&\eqdef \EE\left[\log \frac{q(X,Y)^s}{\sum_{x'\in\Xc}p(x')q(x',Y)^s}\right].
\end{align}
As in the case of matched decoding, this definition can be recovered from the  error exponent,
\begin{equation}
    \Igmis =\frac{d E_0^q(\rho,s) }{d\rho}\Biggr|_{\rho=0} = \lim_{\rho\to 0}\frac{E_0^q(\rho,s)}{\rho}.
\end{equation}

We next see that the generalized mutual information is equal to the BICM capacity of \cite{caire1998bic} when the metric \eqref{eq:bit_metric_sum} is used. Similarly to Section~\ref{section:achievability}, the result does not require the presence of an interleaver of infinite length. Further, the interleaver is actually not necessary for the random coding arguments. 
First, we have the following,
\begin{theorem}\label{theorem:gmiBICM-sum}
The generalized mutual information of the BICM mismatched decoder is equal to the sum of the generalized mutual informations of the independent binary-input parallel channel model of BICM,
\begin{align}
\Igmi
&=\sup_{s>0} ~\sum_{j=1}^m \EE\left[\log\frac{q_j(b_j,Y)^s}{\frac{1}{2}\sum_{b'=0}^1q_j(b_j',Y)^s}\right].
\end{align}
The expectation is carried out according to the joint distribution $p_j(b_j)p_j(y|b_j)$, with $p_j(b_j) = \frac{1}{2}$.
\end{theorem}

\begin{proof}
For a given $s$, and uniform inputs, i.e., $p(x)=\frac{1}{2^m}$, Eq.~\eqref{eq:gmi-s} gives
\begin{align}
\Igmis
& = \EE\left[\log \frac{\prod_{j=1}^m q_j\bigl(b_j(X),Y\bigr)^s}{\sum_{x'}\frac{1}{2^m}\prod_{j=1}^m q_j\bigl(b_j(x'),Y\bigr)^s}\right].
\label{eq:gmi_log}
\end{align}

We now have a closer look at the denominator in the logarithm of \eqref{eq:gmi_log}. The key observation here is that the sum over the constellation points of the product over the binary label positions can be expressed as the product over the label position is the sum of the probabilities of the bits being zero and one, i.e.,
\begin{align}
\sum_{x'}\frac{1}{2^m}\prod_{j=1}^m q_j\bigl(b_j(x'),Y\bigr)^s &=
\frac{1}{2^m} \prod_{j=1}^m \bigl( q_j(b_j'=0,Y)^s + q_j(b_j'=1,Y)^s\bigr)\\
&= \prod_{j=1}^m \Biggl(\frac{1}{2}\bigl(q_j(b_j'=0,Y)^s + q_j(b_j'=1,Y)^s\bigr)\Biggr).
\end{align}
Rearranging terms in \eqref{eq:gmi_log} we obtain,
\begin{align}
\Igmis &= \EE \left[\sum_{j=1}^m \log\frac{q_j\bigl(b_j(x),Y\bigr)^s}{\frac{1}{2}\bigl(q_j(b_j'=0,Y)^s + q_j(b_j'=1,Y)^s\bigr)}\right]\\
&= \sum_{j=1}^m \frac{1}{2} \sum_{b=0}^1 \frac{1}{2^{m-1}}\sum_{x\in\Xjb}\EE\left[ \log\frac{q_j\bigl(b_j(x),Y\bigr)^s}{\frac{1}{2}\sum_{b'=0}^1q_j(b_j',Y)^s}\right],
\end{align}
the expectation expectation being done according to the joint distribution $p_j(b_j)p_j(y|b_j)$, with $p_j(b_j) = \frac{1}{2}$.
\end{proof}

There are a number of interesting particular cases of the above theorem.
\begin{corollary}
For the classical BICM decoder with metric in Eq.~\eqref{eq:bit_metric_sum},
\begin{align}
\Igmi &= \Cbicm{\Xc}. 
\label{eq:gmi_bicm}
\end{align}
\end{corollary}

\begin{proof}
Since the metric $q_j(b_j,y)$ is proportional to $p_j(y|b_j)$, we can identify the quantity
\beq
\EE\left[\log\frac{q_j\bigl(B_j,Y\bigr)^s}{\frac{1}{2}\sum_{b_j'=0}^1 q_j(b_j',Y)^s }\right]
\eeq
as the generalized mutual information of a {\em matched} binary-input channel with transitions $p_j(y|b_j)$. Then, the supremum over $s$ is achieved at $s=1$ and we get the desired result.
\end{proof}

\begin{corollary}
For the max-log metric in Eq.~\eqref{eq:max_metric},
\begin{align}
&\Igmi =\sup_{s>0}\sum_{j=1}^m \EE\left[\log\frac{\bigl(\max_{x\in{\Xc_j^{B}}}p(y|x)\bigr)^s}{\frac{1}{2}\sum_{b=0}^1\bigl(\max_{x'\in{\Xjb}}p(y|x')\bigr)^s}\right].
\label{eq:mi_max_s}
\end{align}
\end{corollary}
Szczecinski {\em et al.} studied the mutual information with this decoder \cite{szczecinski2007dvg}, using this formula for $s = 1$. Clearly, the optimization over $s$ may induce a larger achievable rate, as we see in the next section. More generally, as we shall see later, letting $s=1/(1+\rho)$ in the mismatched error exponent can yield some degradation.

\subsection{BICM with mismatched decoding: numerical results}
\label{section:exponents}

The data-processing inequality for error exponents yields $E^{\rm bicm}_0(\rho,s) \leq E^{\rm cm}_0(\rho)$,
where the quantity in the right-hand side is the coded modulation exponent.
On the other hand, no general relationship holds between $E_0^{\rm ind}(\rho)$ and $E_0^{\rm cm}(\rho)$. As it will be illustrated in the following examples, there are cases for which $E_0^{\rm cm}(\rho)$ can be larger than $E_0^{\rm ind}(\rho)$, and viceversa.

\begin{figure}[htbp]
\begin{center}
\includegraphics[width=0.9\columnwidth]{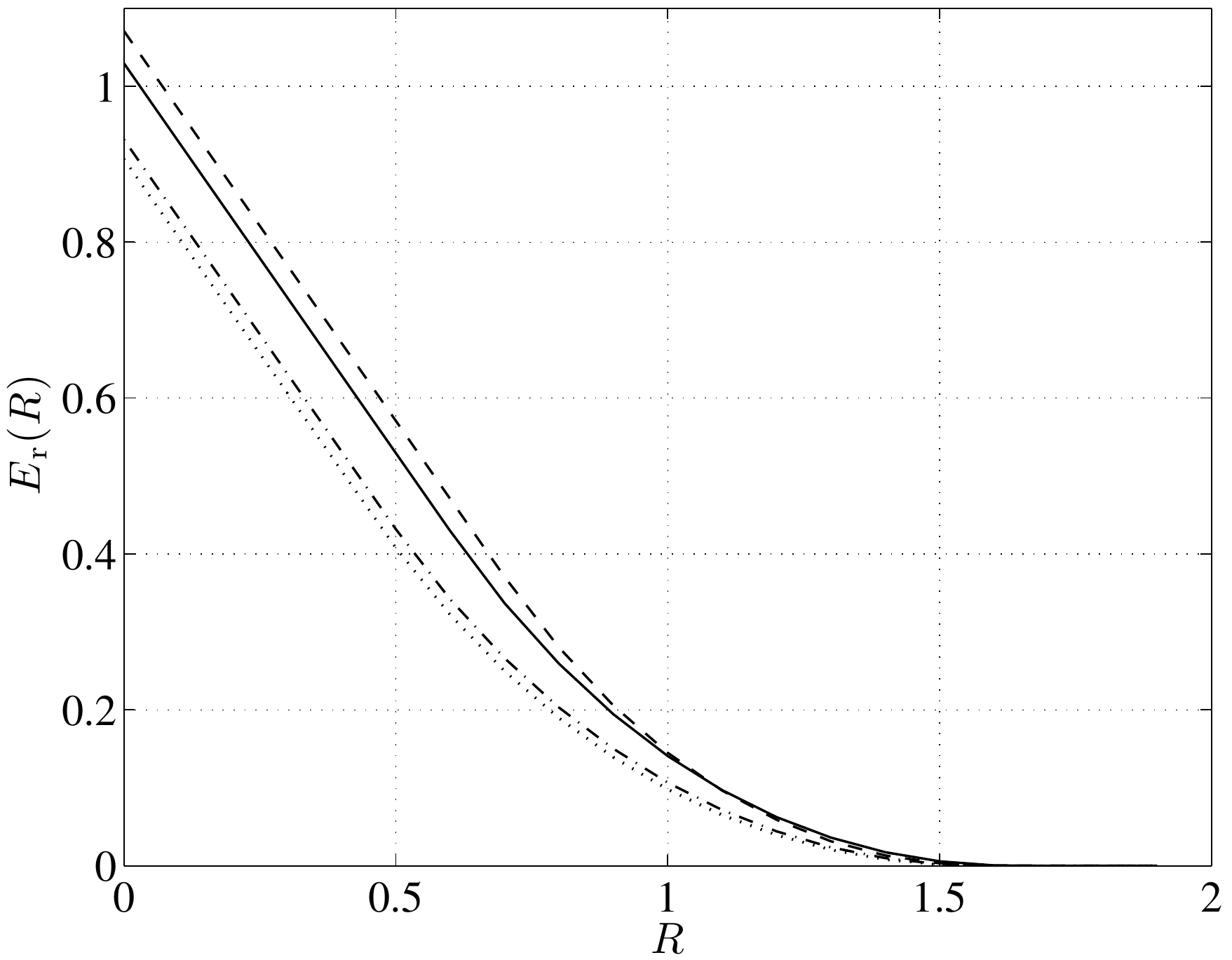}
\caption{Error exponents for coded modulation (solid), BICM with independent parallel channels (dashed), BICM using mismatched metric \eqref{eq:bit_metric_sum} (dash-dotted), and BICM using mismatched metric \eqref{eq:max_metric} (dotted) for $16$-QAM with Gray mapping, Rayleigh fading and $\snr=5$ dB.}
\label{fig:error_exponents_16qamgr_snr5dB_rayleigh}
\end{center}
\vspace{-4mm}
\end{figure}

\begin{figure}[htbp]
\begin{center}
\includegraphics[width=0.9\columnwidth]{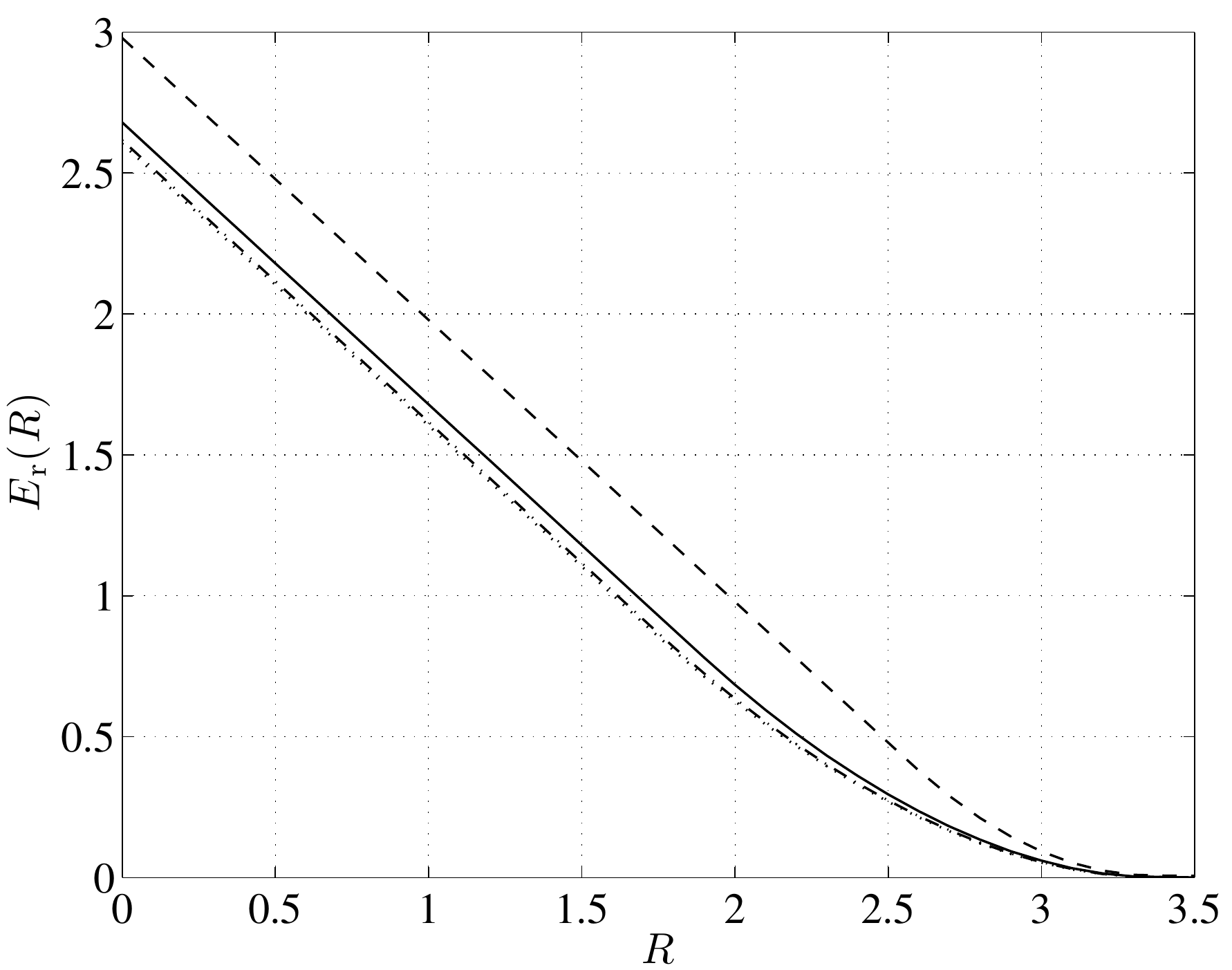}
\caption{Error exponents for coded modulation (solid), BICM with independent parallel channels (dashed), BICM using mismatched metric \eqref{eq:bit_metric_sum} (dash-dotted), and BICM using mismatched metric \eqref{eq:max_metric} (dotted) for $16$-QAM with Gray mapping, Rayleigh fading and $\snr=15$ dB.}
\label{fig:error_exponents_16qamgr_snr15dB_rayleigh}
\end{center}
\vspace{-4mm}
\end{figure}

\begin{figure}[htbp]
\begin{center}
\includegraphics[width=0.9\columnwidth]{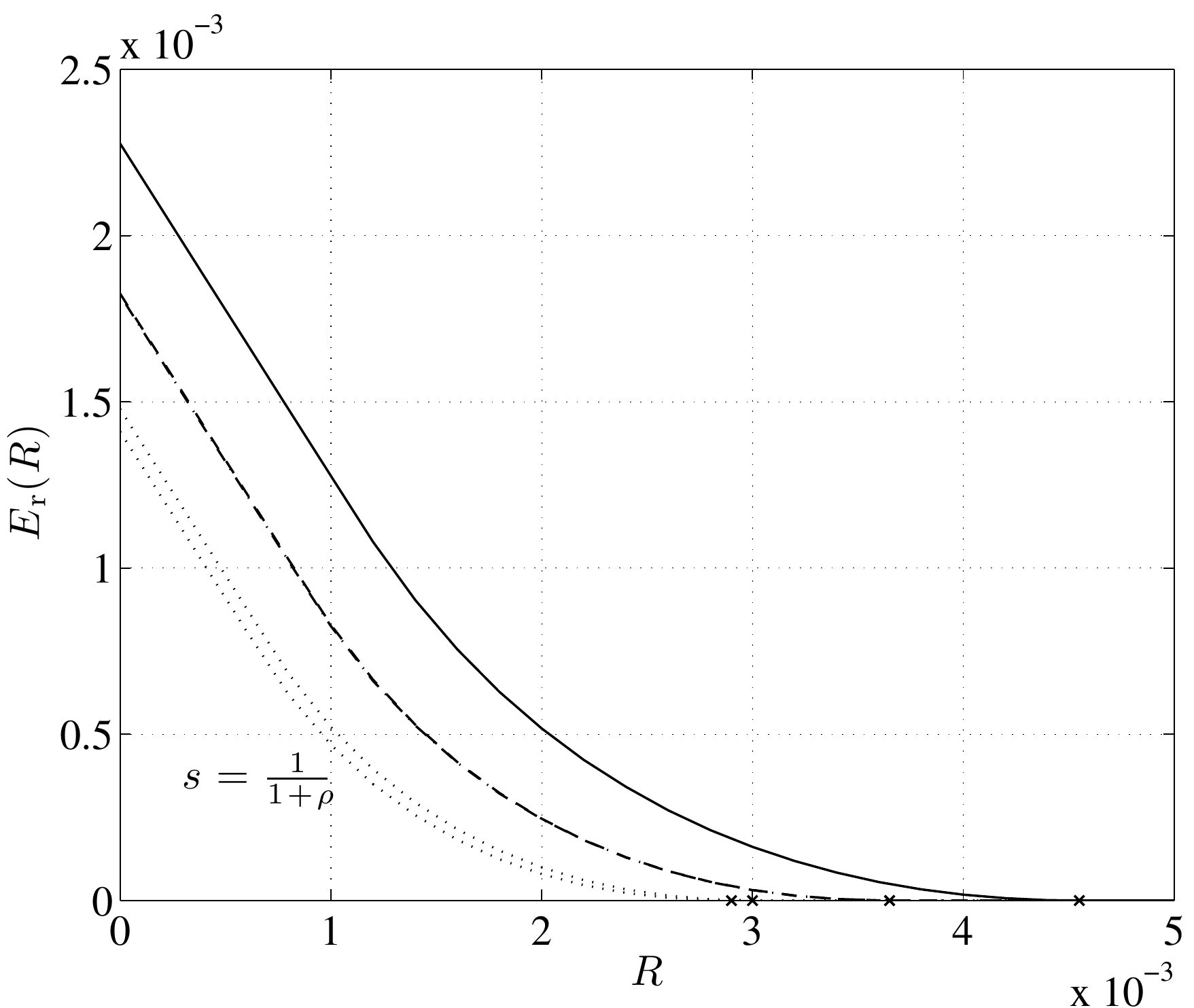}
\caption{Error exponents for coded modulation (solid), BICM with independent parallel channels (dashed), BICM using mismatched metric \eqref{eq:bit_metric_sum} (dash-dotted), and BICM using mismatched metric \eqref{eq:max_metric} (dotted) for $16$-QAM with Gray mapping, Rayleigh fading and $\snr=-25$ dB. Crosses correspond to (from right to left) coded modulation, BICM with metric \eqref{eq:bit_metric_sum}, BICM with metric \eqref{eq:max_metric} and BICM with metric \eqref{eq:max_metric}  with $s=1$.}
\label{fig:error_exponents_16qamgr_snr_25dB_rayleigh}
\end{center}
\vspace{-4mm}
\end{figure}

Figures \ref{fig:error_exponents_16qamgr_snr5dB_rayleigh}, \ref{fig:error_exponents_16qamgr_snr15dB_rayleigh} and \ref{fig:error_exponents_16qamgr_snr_25dB_rayleigh} show the error exponents for coded modulation (solid), BICM with independent parallel channels (dashed), BICM using mismatched metric \eqref{eq:bit_metric_sum} (dash-dotted), and BICM using mismatched metric \eqref{eq:max_metric}  (dotted) for $16$-QAM with Gray mapping, Rayleigh fading and $\snr=5,15, -25$ dB, respectively.  Dotted lines labeled with $s=\frac{1}{1+\rho}$ correspond to the error exponent of BICM using mismatched metric \eqref{eq:max_metric} letting $s=\frac{1}{1+\rho}$.
The parallel-channel model gives a larger exponent than the coded modulation, in agreement with the cutoff rate results of \cite{caire1998bic}. In contrast, the mismatched-decoding analysis yields a lower exponent than coded modulation. As mentioned in the previous section, both BICM models yield the same capacity.

In most cases, BICM with a max-log metric \eqref{eq:max_metric} incurs in a marginal loss in the exponent for mid-to-large SNR. In this SNR range, the optimized exponent and that with $s=\frac{1}{1+\rho}$ are almost equal. For low SNR, the parallel-channel model and the mismatched-metric model with \eqref{eq:bit_metric_sum} have the same exponent, while we observe a larger penalty when metrics \eqref{eq:max_metric} are used. As we observe, some penalty is incurred at low SNR for not optimizing over $s$. We denote with crosses the corresponding achievable information rates.

\begin{figure}[htbp]
\begin{center}
\includegraphics[width=0.9\columnwidth]{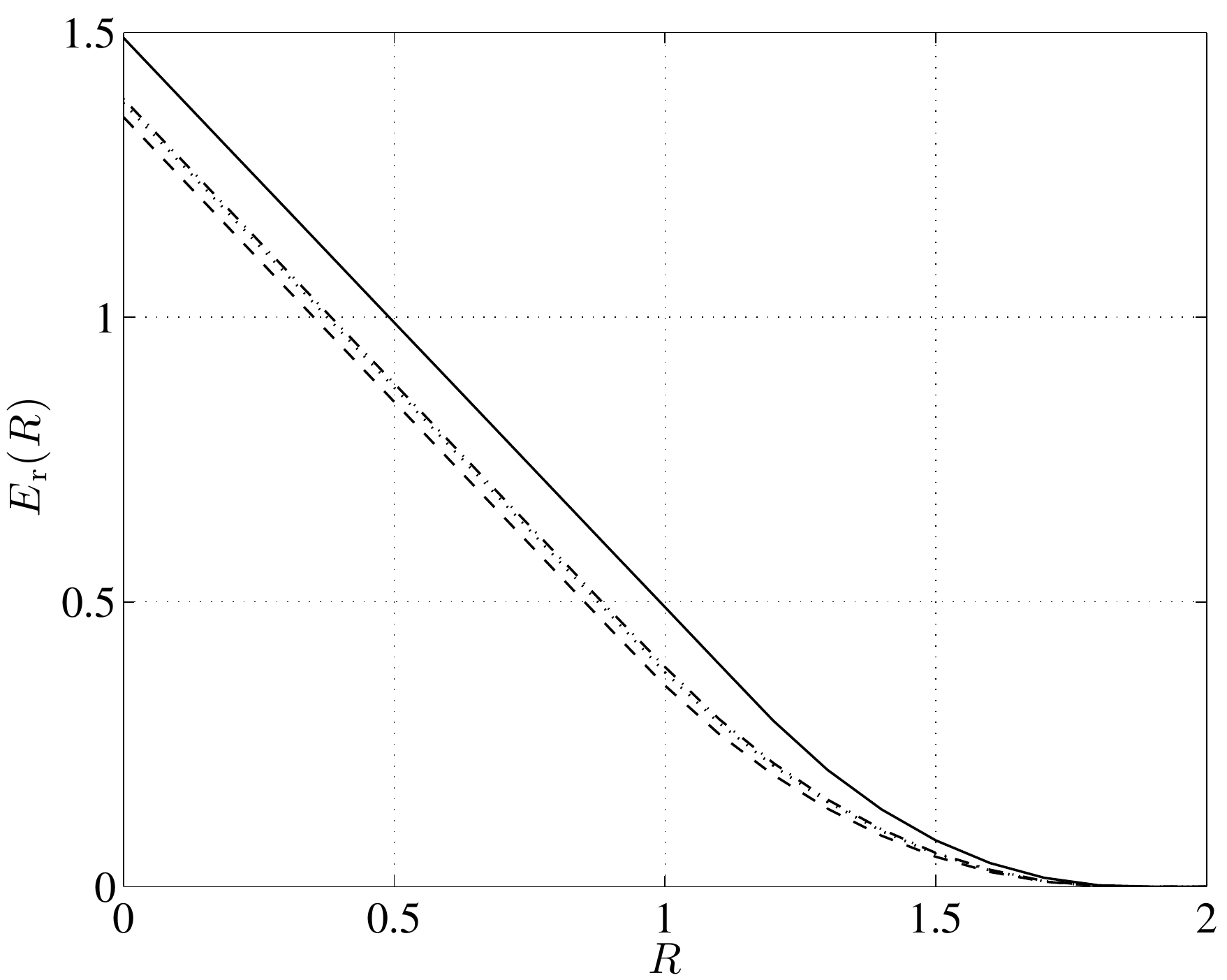}
\caption{Error exponents for coded modulation (solid), BICM with independent parallel channels (dashed), BICM using mismatched metric \eqref{eq:bit_metric_sum} (dash-dotted), and BICM using mismatched metric \eqref{eq:max_metric} (dotted) for $8$-PSK with Gray mapping, AWGN and $\snr=5$ dB.}
\label{fig:error_exponents_8pskgr_snr5dB_awgn}
\end{center}
\vspace{-4mm}
\end{figure}

An interesting question is whether the error exponent of the parallel-channel model is always larger than that of the mismatched decoding model. The answer is negative, as illustrated in Figure \ref{fig:error_exponents_8pskgr_snr5dB_awgn}, which shows the error exponents for coded modulation (solid), BICM with independent parallel channels (dashed), BICM using mismatched metric \eqref{eq:bit_metric_sum} (dash-dotted), and BICM using mismatched metric \eqref{eq:max_metric} (dotted) for $8$-PSK with Gray mapping in the unfaded AWGN channel.

\section{Extrinsic Side Information}
\label{section:feedback}

Next to the classical decoder described in Section~\ref{section:model}, iterative decoders have also received much attention \cite{li1998bic,li1999tcm,tenbrink2000did,tenbrink1998idq,tenbrink1999cid} due to their improved performance. Iterative decoders can also be modelled as mismatched decoders, where the bit decoding metric is now of the form
\beq\label{eq:bit_metric_bicm-fb}
q_j(b,y) = \sum_{x'\in\Xjb} p(y|x') \prod_{j'\neq j}\prior_{j'}\bigl(b_{j'}(x')\bigr),
\eeq
where we denote by $\prior_j(b)$ the extrinsic information, i.e., the ``a priori'' probability that the $j$-th bit takes the value $b$. Extrinsic information is commonly generated by the decoder of the binary code $\Cc$.
 Clearly, we have that $\prior_j(0)+\prior_j(1) = 1$, and $0\leq \prior_j(0),\prior_j(1)\leq 1$. Without extrinsic information, we take $\prior_j(0) = \prior_j(1) = \frac{1}{2}$, and the metric is given by Eq.~\eqref{eq:bit_metric_sum}.

In the analysis of iterative decoding, extrinsic information is often modeled as a set of random variables $\Prior_j(0)$, where we have defined without loss of generality the variables with respect to the all-zero symbol. We denote the joint density function by $p(\prior_1(0),\dotsc,\prior_m(0))=\prod_{j=1}^mp(\prior_j(0))$. We discuss later how to map the actual extrinsic information generated in the decoding process onto this density. The mismatched decoding error exponent $E_0^q(\rho,s)$ for metric \eqref{eq:bit_metric_bicm-fb} is given by Eq.~\eqref{eq:E0bicm}, where the expectation is now carried out according to the joint density $p(x)p(y|x)p(\prior_1(0))\dotsi p(\prior_m(0))$. Similarly,  the generalized mutual information is again obtained as $\Igmi = \max_s \lim_{\rho\to 0}\frac{E_0^q(\rho,s)}{\rho}$ . 

It is often assumed \cite{tenbrink2000did} that the decoding metric acquires knowledge on the symbol $x$ effectively transmitted, in the sense that for any symbol $x'\in\Xc$, the $j$-th bit decoding metric is given by
\begin{align}\label{eq:bit_metric_bicm-fb-a}
q_j\bigl(b_j(x')=b,y\bigr) &= \sum_{x''\in\Xjb} p(y|x'') \prod_{j'\neq j}\prior_{j'}\bigl(b_{j'}(x'')\oplus b_{j'}(x)\bigr),
\end{align}
where $\oplus$ denotes the binary  addition. Observe that extrinsic information is defined relative to the transmitted symbol $x$, rather than relative to the all-zero symbol. If the $j$-th bit of the symbols $x''$ and $x$ coincide, the extrinsic information for bit zero $\prior_j(0)$ is selected, otherwise the extrinsic information $\prior_j(1)$ is used.

For the metric in Eq.~\eqref{eq:bit_metric_bicm-fb-a}, the proof presented in Section~\ref{sec:dataProcessing} of the data processing inequality fails because the integrand in Eq.~\eqref{eq:inside_log_hoelder} cannot be decomposed into a product of separate terms respectively depending on $x$ and $x'$, the reason being that the metric $q(x',y)$ varies with $x$.

On the other hand, since the symbol metric $q(x',y)$ is the same for all symbols $x'$, the decomposition of the generalized mutual information as a sum of generalized mutual informations across the $m$ bit labels in Theorem~\ref{theorem:gmiBICM-sum} remains valid, and we have therefore
\begin{align}
    \Igmi=\sum_{j=1}^m \EE\left[\log\frac{q_j(B_j,Y)}{\frac{1}{2}\sum_{b'=0}^1q_j(b_j=b',Y)}\right].
\end{align}
This expectation is carried out according to $p(b_j)p_j(y|b_j)p(\prior_1(0))\dotsi p(\prior_m(0))$, with $p(b_j) = \frac{1}{2}$.
Each of the summands can be interpreted as the mutual information achieved by non-uniform signalling in the constellation set $\Xc$, where the probabilities according to which the symbols are drawn are a function of the extrinsic informations $\prior_j(\cdot)$.
The value of $\Igmi$ may exceed the channel capacity \cite{tenbrink2000did}, so this quantity is a pseudo-generalized mutual information, with the same functional form but lacking operational meaning as an achievable rate by the decoder.

Alternatively, the metric in Eq.~\eqref{eq:bit_metric_bicm-fb} may depend on the hypothesized symbol $x'$, that is
\begin{align}\label{eq:bit_metric_bicm-fb-b}
q_j\bigl(b_j(x')=b,y\bigr) &= \sum_{x''\in\Xjb} p(y|x'') \prod_{j'\neq j}\prior_{j'}\bigl(b_{j'}(x'')\oplus b_{j'}(x')\bigr).
\end{align}
Differently from Eq.~\eqref{eq:bit_metric_bicm-fb-a}, the bit metric varies with the hypothesized symbol $x'$ and not with the transmitted symbol $x$. Therefore, Theorem~\ref{theorem:gmiBICM-sum} cannot be applied and the generalized mutual information cannot be expressed as a sum of mutual informations across the bit labels. On the other hand, the data processing inequality holds and, in particular, the error exponent and the generalized mutual information are upper bounded by that of coded modulation. Moreover, we have the following result.
\begin{theorem}
In the presence of perfect extrinsic side information, the error exponent with metric \eqref{eq:bit_metric_bicm-fb-b} coincides with that of coded modulation.
\end{theorem}
\begin{proof}
With perfect extrinsic side information, all the bits $j'\neq j$ are known, and then
\beq
\prod_{j'\neq j}\prior_{j'}\bigl(b_{j'}(x'')\oplus b_{j'}(x')\bigr) = 
\begin{cases}
1 &\text{when $x'' = x'$,}\\
0 &\text{otherwise,}
\end{cases}
\eeq
which guarantees that only the symbol $x'' = x'$ is selected. Then,
$
q_j\bigl(b_j(x')=b,y\bigr) = p(y|x')
$
and the symbol metric becomes $q(x',y) = p(y|x')^m$ for all $x'\in\Xc$. As we showed in Eq.~\eqref{eq:sufficiencyHoelder}, this is precisely the condition under which the error exponent (and the capacity) with mismatched decoding coincides that of coded modulation.
\end{proof}

The above result suggests that with perfect extrinsic side information, the gap between the error exponent (and mutual information) of BICM and that of coded modulation can be closed if one could provide perfect side information to the decoder.
A direct consequence of this result is that the generalized mutual information with BICM metric \eqref{eq:bit_metric_bicm-fb} and perfect extrinsic side information is equal to the mutual information of coded modulation. An indirect consequence of this result is that the multi-stage decoding \cite{imai1977nmc,wachsmann1999mct} does not attain the exponent of coded modulation, even though its corresponding achievable rate is the same. The reason is that the decoding metric is not of the form $c\, p(y|x)^s$, for some constant $c$ and $s$, except for the last bit in the decoding sequence.
We hasten to remark that the above rate in presence of perfect extrinsic side information need not be achievable, in the sense that there may not exist a mechanism for accurately feeding the quantities $\prior_j(b)$ to the demapper. Moreover, the actual link to the iterative decoding process is open for future research.

\section{Conclusions}
\label{section:conclusions}

We have presented a mismatched-decoding analysis of BICM, which is valid for arbitrary finite-length interleavers. We have proved that the corresponding generalized mutual information coincides with the BICM capacity originally given by Caire {\em et al.} modeling BICM as a set of independent parallel channels. More generally, we have seen that the error exponent cannot be larger than that of coded modulation, contrary to the analysis of BICM as a set of parallel channels.
For Gaussian channels with binary reflected Gray mapping, the gap between the BICM and CM error exponents is small, as found by Caire {\em et al.} for the capacity.
We have also seen that the mutual information appearing in the analysis of iterative decoding of BICM via EXIT charts admits a representation as a form of generalized mutual information. However, since this quantity may exceed the capacity, its operational meaning as an achievable rate is unclear. We have modified the extrinsic side information available to the decoder, to make it dependent on the hypothesized symbol rather than on the transmitted one, and shown that the corresponding error exponent is always lower bounded by that of coded modulation. In presence of perfect extrisinc side information, both error exponents coincide. The actual link to the iterative decoding process is open for future research.

\newpage
\bibliographystyle{IEEE}

\newpage

\newpage

\newpage

\newpage

\newpage

\end{document}